\documentclass{article}
\usepackage{blindtext}
\usepackage[a4paper]{geometry}
\usepackage{graphicx}
\usepackage{bm}
\usepackage{amsmath}
\usepackage{amsfonts}
\usepackage{amssymb}

\title{Network Model with Application to Allergy Diseases}
\author{Konrad Furmańczyk$^{1,2}$, Wojciech Niemiro$^{3,4}$, Mariola Chrzanowska$^{5,2}$, \\Marta Zalewska$^{2}$}

\begin{document}

\maketitle 

\begin{center}
	\small  
	Department of Applied Mathematics, Institute of Information Technology, \\Warsaw University of Life Sciences, Poland, e-mail correspondence: konrad\_furmanczyk@sggw.edu.pl $^1$\\
	
	Department of Prevention of Environmental Hazards, Allergology and Immunology, \\Medical University of Warsaw, Poland $^2$\\
	
	Faculty of Mathematics, Informatics and Mechanics
	University of Warsaw, Poland $^3$\\
	
	Faculty of Mathematics and Computer Science
	Nicolaus Copernicus University, Torun, Poland $^4$\\
	
	Department of Statistics and Econometrics, Institute of Economics and Finance, \\Warsaw University of Life Sciences, Poland $^5$\\
\end{center}

\begin{abstract}
We propose a new graphical model to describe the comorbidity of allergic diseases. We present our model in two versions. First, we introduce a generative model that correctly reflects the variables' causal relationship. Then we propose an approximation of the generative model by another misspecified model that is computationally more efficient and easily interpretable. We will focus on the misspecified version, which we consider more practical. We include in the model two directed graphs, one graph of known dependency between the main binary variables (diseases), and a second graph of the dependence between the occurrence of the diseases and their symptoms. In the model, we also consider additional auxiliary variables. The proposed model is evaluated on a cross-sectional multicentre study in Poland on the ECAP database (www.ecap.pl). An assessment of the stability of the proposed model was obtained using bootstrap and jackknife techniques.
\end{abstract}



\section{Introduction}
Modeling dependency between different binary variables is an essential statistical task with many applications in medicine, life sciences, economics, and sociology. The basic statistical tools used in such modeling are the autologistic model (\cite{b1}) and network modeling based on the Ising model model (\cite{b2}, \cite{b3}, \cite{b3a}, \cite{b3b}, \cite{b3c}, \cite{b3d}), 
\\which is a special case of autologistic model. More general information of graphical models for discrete data can be found in \cite{b4} and \cite{hbook}. The classical autologistic model (\cite{b1}) has been applied many times, e.g., in epidemiology, marketing, agriculture, ecology, forestry, geography, and image analysis (\cite{b2a}, \cite{b5}, \cite{b9}, \cite{b15a}, \cite{b19a}).
\\ \cite{b5} considered a centered autologistic model with more interpretable parameters which describe spatial dependence. \cite{b6} propose a binary vector autologistic regressive model in time and use regularization methods to estimate a sparse network. Most common approach for estimation of the model parameters is pseudo‐likelihood (\cite{b7}) estimation. \cite{b8} described maximum likelihood via MCMC and recommended a heuristic method of estimation (averaging estimators).
\\Recently in the medical area, \cite{b9} invented and applied autologistic network model for a disease progression study. Their model work with complex spatial and temporal dependencies in muscle strength among different muscles. 
\cite{b9} use pseudo‐likelihood (\cite{b7}) to estimate the model parameters. To overcome a large number of pairwise spatial associations, they apply the least absolute shrinkage and selection operator (LASSO) (\cite{b10}). 
\\This paper proposes a new graphical model related to but different from the autologistic model. Our model aims to describe the interdependence of allergic diseases in contrast to most studies that do not consider dependencies between allergies (\cite{b17a}, \cite{b17b}, \cite{b17c}).
\\We present our model in two versions. First, we introduce a generative model that correctly reflects the variables' causal relationship. Then we propose an approximation of the generative model by another misspecified model that is computationally more efficient and easily interpretable. We will focus on the misspecified version, which we consider more practical. In both versions of our model, we will consider typical allergic disease symptoms, family history of allergic disease, and control variables as covariate variables. We describe information about the coexistence of certain allergic diseases (binary variables) by a directed acyclic graph (DAG), which will allow us to estimate only those model parameters responsible for the strength of the relationship between individual allergic diseases. The second graph will describe the relationship between particular symptoms and the occurrence of these diseases. In the generative model, the edges lead from diseases to symptoms, corresponding to causal relations. In the misspecified model, we reverse the direction of edges: they lead from symptoms to diseases. Additionally, we consider the potential impact of a family history of allergy on the occurrence of the disease. The applied approach based on the graphs of known interdependencies of binary variables is very flexible. It will allow us to estimate only those parameters for which experts have knowledge about the presence of interdependencies. This approach will avoid the inevitable collinearity between the variables under consideration and significantly reduce computational costs. Our model was naturally divided into separate logistic models for individual allergy diseases. Each individual logistic regression is estimated by standard glm procedure and also by weighted logistic estimation (\cite{b11}). This is because we study the so-called rare diseases. The dataset is divided into a learning and testing sample to recommend which estimation method is appropriate, and the ROC curve and average AUC on the testing sample are determined from 20 repetitions via the bootstrap and jackknife method. This approach is general and can be applied to any binary variable dependency model. It can also be extended to the high-dimensional case by adding the Lasso penalty to the conditional likelihood estimation.
\\The network considered in our study has a relatively small size. Therefore we can compare two versions of our model. For five (different) scenarios of covariates, we compute the ’diagnostic’ probability of diseases on given symptoms for both the generative and the misspecified models. The obtained differences are negligible. We can treat the misspecified model as a good approximation for the more logically consistent but computationally expensive generative model. The rest of the paper is organized as follows. In Section 2, we introduce the proposed model with estimation. Section 3 describes the construction of the proposed model to a real big epidemiological data set with estimation results for two methods (standard logistic estimation and weighted logistic estimation), and we compare the generative model with the misspecified model. At the end of this section, we present the evaluation of the proposed model. Section 4 contains a discussion of our approach, and in Section 5, we conclude our work.

\section{Network Model}

\subsection{Genarative model}
Our proposed model contains four groups of variables. In the first group, we consider a random vector $\mathbf{Y}=(Y_1,\ldots, Y_p)^T $  with binary components.These variables will determine a patient's presence or absence of a given allergic disease. In our application we describe the interdependencies of $p$ most common allergic diseases. Taking into account the known co-occurrence of diseases, these relationships can be described by a directed graph with the adjacency matrix $\mathbf{A}=(a_{ki})$ as follows: $a_{ki}=1$ if $Y_i$ is affected by $Y_k$ and otherwise $a_{ki}=0.$ 
The random variables $Y_i$ for $i=1,\ldots,p$ are considered as vertices of a graph, where edge $(k,i)$ occurs when $Y_k$ affects $Y_i$. 
\\In the second group, we have a random vector of symptoms of our diseases $\mathbf{S}=(S_1,\ldots,S_m)^T$. The remaining two groups consist of common factors $\mathbf{F}=(F_1,\ldots, F_l)^T,$ which can affect all considered diseases, and a vector of additional covariates $\mathbf{X}=(X_1,\ldots,X_r)^T$ such as sex, age, region of a patient, etc. 
For example, genetic features can be considered as common factors for allergic diseases. Symptoms $S_i$ can be continuous or discrete random variables. It is usually known which symptoms of diseases are characteristic for each disease. This knowledge can be represented in a similar way as in the case of correlations among diseases by an directed graph with adjacency matrix $\mathbf{B}=(b_{kj})$ such that:  $b_{kj}=1$ if  $Y_k$ causes $S_j$ and otherwise $b_{kj}=0.$      
\\Our generative model is therefore a graphical model that includes diseases $\mathbf{Y}$, symptoms $\mathbf{S}$, common factors $\mathbf{F}$ and additional covariates $\mathbf{X}$. 
The structure of this graph is described by edges among $\mathbf{Y, S}$ variables given by matrices $\mathbf{A, B}$, and all edges leading from $\mathbf{F, X}$ variables to all components of $\mathbf{Y, S}$. We assume that the graph corresponding to the adjacency matrix $\mathbf{A}$ is acyclic. Consequently, the whole consider graph is a directed acyclic graph (DAG). 
The conditional probability distribution of $\mathbf{Y,S}$ is given by 
\begin{equation}
	\label{ff1} 
	\begin{split}
	P(\mathbf{Y}=\mathbf{y},\mathbf{S}=\mathbf{s}| \mathbf{F}=\mathbf{f},\mathbf{X}=\mathbf{x})&=\prod_{i=1}^{p}P(Y_{i}=y_{i}|\mathbf{Y}_{pa}(Y_i),\mathbf{F}=\mathbf{f},\mathbf{X}=\mathbf{x}) \\ 
	&\times 
 \prod_{j=1}^{m}P(S_{j}=s_{j}|\mathbf{Y}_{pa}(S_j),\mathbf{F}=\mathbf{f},\mathbf{X}=\mathbf{x}),
	\end{split}
\end{equation}
where $\mathbf{Y}_{pa}(Y_i)=\{Y_k: Y_k \to Y_i \}$ is a set of diseases which affect the occurrence of disease $Y_i$,  $\mathbf{Y}_{pa}(S_j)=\{Y_k: Y_k \to S_j \} $ is a set of diseases which cause symptom  $S_j$.
   We assume the following parametric form of conditional distribution:
\begin{equation}
	\label{ff2} 
\log\left( \frac{P(Y_i=1|\mathbf{Y}_{pa}(Y_i),\mathbf{F}=\mathbf{f},\mathbf{X}=\mathbf{x})}{P(Y_i=0|\mathbf{Y}_{pa}(Y_i),\mathbf{F}=\mathbf{f},\mathbf{X}=\mathbf{x})}\right)
=\omega_{0i}+\sum_{k=1}^{p}a_{ki}\omega_{ki}Y_k+	 \mathbf{x}^T\bm{\alpha}_i+\mathbf{f}^T\mathbf{\bm{\beta}}_i, 
\end{equation}   
 \begin{equation}
 	\label{ff3} 
 	\log\left( \frac{P(S_j=1|\mathbf{Y}_{pa}(S_j),\mathbf{F}=\mathbf{f},\mathbf{X}=\mathbf{x})}{P(S_j=0|\mathbf{Y}_{pa}(S_j),\mathbf{F}=\mathbf{f},\mathbf{X}=\mathbf{x})}\right)
 	=\gamma_{0j}+\sum_{k=1}^{p}b_{kj}\gamma_{kj}Y_k+	 \mathbf{x}^T\bm{\delta}_j+\mathbf{f}^T\mathbf{\bm{\epsilon}}_j. 
 \end{equation}     
   where model parameters: $\omega_{0i}\in R, \omega_{ki}\in R, \bm{\alpha}_i \in R^r, \bm{\beta}_i\in R^l,\gamma_{0j}\in R, \gamma_{kj} \in R, \bm{\delta}_{j} \in R^r, \bm{\epsilon}_{j} \in R^l.$ 
 \\Since the conditional probability (\ref{ff1}) consists of the product of $p+m$ probabilities (factors), the parameters of each factor can be estimated separately by fitting a standard logistic regression procedure. 
 \subsection{Misspecified model}
 Unfortunately, the model presented in the previous subsection is computationally demanding, and moreover, its parameters are not easy to interpret. We propose using another model that does not reflect causal relations between variables correctly but is computationally easier in a big network and has parameters with simple, intuitive meanings. We say that this model is misspecified. We change the direction of edges joining symptoms and diseases. Entries of adjacency matrix $\mathbf{B}$ will now be interpreted as follows: $b_{ij}=1$ indicates the presence of arrow $Y_i \leftarrow S_j$. We assume that the remaining edges of the graph are the same as in generative model. In the misspecified model, equation (\ref{ff1}) is replaced by equation (\ref{ff4}), and equations (\ref{ff2})-(\ref{ff3}) are replaced by equation (\ref{ff5}) as follows:
\begin{equation}
	\label{ff4} 
	\begin{split}
		P(\mathbf{Y}=\mathbf{y}|\mathbf{S}=\mathbf{s}, \mathbf{F}=\mathbf{f},\mathbf{X}=\mathbf{x})&=\prod_{i=1}^{p}P(Y_{i}=y_{i}|\mathbf{Y}_{pa}(Y_i),\mathbf{S}_{pa}(Y_i),\mathbf{F}=\mathbf{f},\mathbf{X}=\mathbf{x}),
	\end{split}
\end{equation}
where $\mathbf{S}_{pa}(Y_i)=\{S_j: Y_i \leftarrow S_j \}$ is a set of symptoms related to occurrence of disease $Y_i$. Similarly, as in generative model, we assume a log-linear form of conditional distributions.
To simplify notation, we use the same symbols for the parameters for both models.  
\begin{equation}
\begin{split}
	\label{ff5} 
	\log\left( \frac{P(Y_i=1|\mathbf{Y}_{pa}(Y_i),\mathbf{S}_{pa}(Y_i),\mathbf{F}=\mathbf{f},\mathbf{X}=\mathbf{x})}{P(Y_i=0|\mathbf{Y}_{pa}(Y_i),\mathbf{S}_{pa}(Y_i),\mathbf{F}=\mathbf{f},\mathbf{X}=\mathbf{x})}\right)
	&=\omega_{0i}+\sum_{k=1}^{p}a_{ki}\omega_{ki}Y_k\\&+\sum_{j=1}^{m}b_{ij}\gamma_{ij}S_j+	 \mathbf{x}^T\bm{\alpha}_i+\mathbf{f}^T\mathbf{\bm{\beta}}_i. 
\end{split}
\end{equation}   
 \\We do not always have appropriate sample sizes for rare diseases and work with imbalanced datasets. In such cases, we may improve prediction accuracy for logistic regression using weighted logistic regression (\cite{b11}, \cite{b12}) or apply a machine learning algorithm such as use SMOTE Simple Genetic Algorithm (\cite{b15}) to determine the sampling rate of each example in order to get unequal synthetic samples or using undersampling or oversampling (\cite{b13}). 
However, resampling techniques do not easily transfer to dependent logistic regression equations. For this reason, in the paper, we use weighted regression as in \cite{b11}. Following the approach of \cite{b11} we penalized misclassification costs of events and non-events differently by penalty weights $w_1$ and $w_0$ in the log-likelihood function for each $i$ equation
$$min_{\bm{\theta}_i} \left\lbrace -w_1\sum_{j=1}^{n}y_{ij} log(\sigma(\mathbf{z}_j^T\bm{\theta}_i))-w_0\sum_{j=1}^{n}(1-y_{ij}) log(1-\sigma(\mathbf{z}_j^T\bm{\theta}_i))\right\rbrace  ,$$ where $n$ is a sample size, $w_1=\frac{\tau_i}{\bar{y_i}}$ and $w_0=\frac{1-\tau_i}{1-\bar{y_i}}$, and $\tau_i$ denoting the population fraction of events induced by choice-based sampling and $\bar{y_i}$ denoting the sample proportion of events, $\bm{\theta}_i$ is a vector of all parameters, $\mathbf{z}_j$ is a vector of all predictors, and $\sigma(x)=\frac{exp(x)}{1+exp(x)}$.

\section{A Model of Allergic Diseases}

The proposed model of disease interdependence will be used to investigate the prevalence of allergic diseases in Poland based on a big epidemiological study ECAP (\cite{b14}). The study was conducted according to the guidelines of the Declaration of Helsinki, and approved by the Ethics Committee of the Medical University of Warsaw, Poland (KB/206/2005).
The study method involved the use of questionnaires adapted for Central and Eastern Europe based on the European Community Respiratory Health Survey II (ECRHS II) (\cite{b15}) and International Study of Asthma and Allergies in Childhood (ISAAC) (\cite{b16}), which had been used as part of a larger project, titled the Implementation of a System for the Prevention and Early Detection of Allergic Diseases in Poland (\cite{b17}). The project was conducted in eight urban areas and one rural area. The study had two stages; the first stage involved grouping the 22,500 respondents based on their questionnaire responses using of a Personal Digital Assistant (PDA); the second stage involved complementary examination (4,783 patients) of a subgroup of stage I respondents who underwent a medical examination. The final data set contains 18,617 units (cases, response) and 1,225 variables (mostly binary).
\subsection{The structure of the model}

\begin{figure}
	\includegraphics[width=9cm, height=9cm]{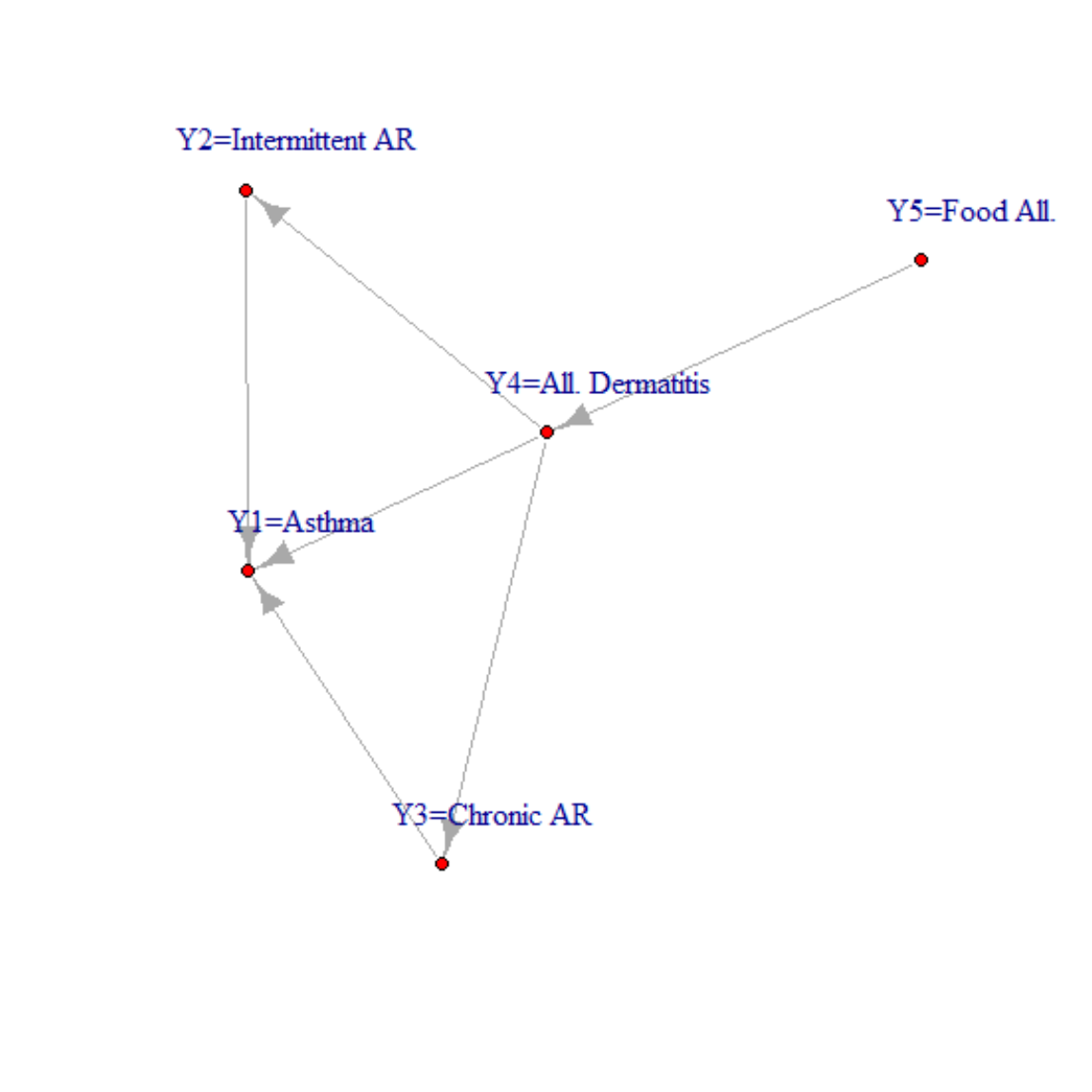}
	\caption{\label{Fig1} Graph with adjacency matrix $\mathbf{A}$.}
\end{figure}

In the first group for the interdependence study, we selected the following diseases, allergic: $Y_1$ -atopic asthma, $Y_2$ -intermittent allergic rhinitis, $Y_3$ -chronic allergic rhinitis, $Y_4$ -allergic dermatitis, $Y_5$ -food allergy.
Figure \ref{Fig1} illustrates the assumed dependencies between these allergic diseases.
The structure of this graph is based on the expert knowledge taken from the medical literature and on discussion with medical doctors (\cite{b18}, \cite{b19}).

\begin{figure}
	\includegraphics[width=9cm, height=9cm]{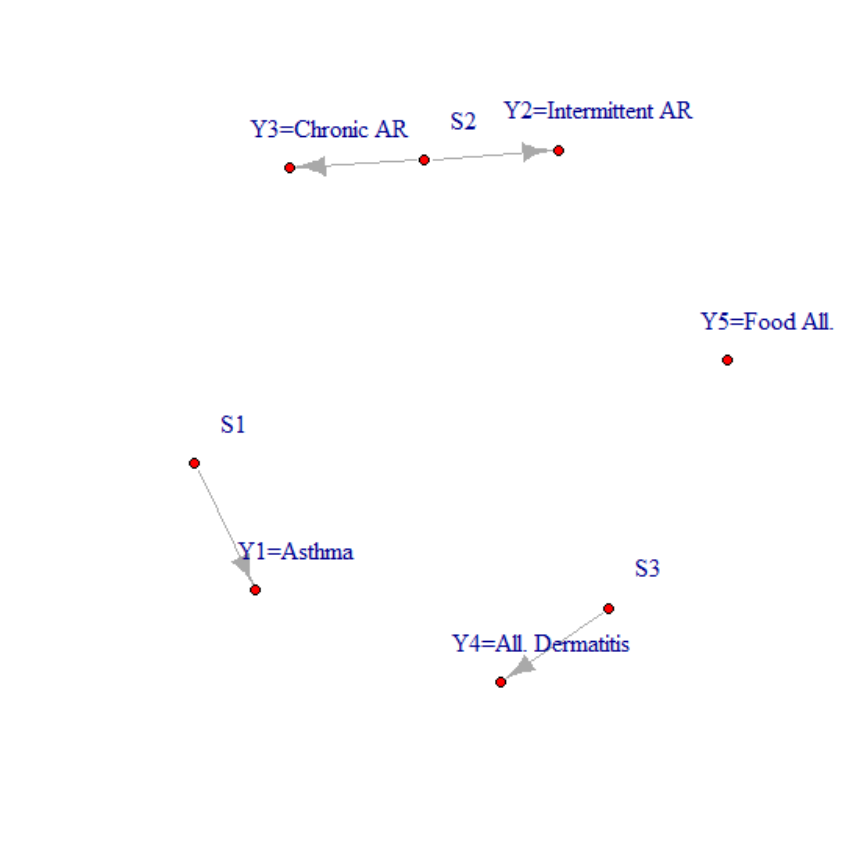}
	\caption{\label{Fig2} Graph with adjacency matrix $\mathbf{B}$ (direction of arrows is as the misspecified model).}
\end{figure}

 In the second group, we consider typical symptoms of those allergic diseases: $S_1$ -Have you had wheezing or whistling in your chest at any time in the last 12 months?; $S_2$ - Have you ever had a problem with sneezing or a runny or blocked nose when you did not have fever, a cold, or the flu?; $S_3$ -Have you ever had eczema or any other form of skin allergy? Additionally we consider in this group history of allergy diseases in the family as a common factors: $F_1$-Does anyone in your immediate family suffer from allergies? - mother; $F_2$-Does anyone in your immediate family suffer from allergies? - father; $F_3$-Does anyone in your immediate family suffer from allergies? -  siblings of the child being tested; $F_4$-Does anyone in your immediate family suffer from allergies? -grandparents on mother's side; $F_5$-Does anyone in your immediate family suffer from allergies? - grandparents on father's side.
Figure \ref{Fig2} shows a relationship graph of the considered allergic diseases and their typical symptoms. This presence of the edges in this graph is also based on expert knowledge (see remarks concerning Figure \ref{Fig1}). Note that the direction of arrows lead from symptoms to diseases which corresponds to the misspecified model.

 In the last group, we consider control covariates for respondents such as: $X_1$ age of patients with three age group: children 6-7 y.o., children 13-14 y.o., and adults (20-44 y.o.). This variable we replaced by new two binary variables: $X_1$ for children 13-14 y.o. and $X_2$ for adults., $X_3$ binary variables with 1 for urban area, $X_4$ -sex (binary variable, 1 for male).

\subsection{Generative and misspecified models of allergy diseases}
We recall that the generative model is a Bayesian Network in which diseases cause symptoms. Conditionally on covariates $\mathbf{F}$ and $\mathbf{X}$ the joint probability distribution of $(\mathbf{Y},\mathbf{S})$ is determined by the set of conditional probabilities: 
$$P(\mathbf{Y},\mathbf{S})=P(Y_1|Y_2,Y_3,Y_4)P(Y_2|Y_4)P(Y_3|Y_4)P(Y_4|Y_5)P(Y_5)P(S_1|Y_1)P(S_2|Y_2,Y_3)P(S_3|Y_4).$$
(We omitted $\mathbf{F}$ and $\mathbf{X}$ in this formula).
For all conditional probabilities, we assume the logistic form of those probabilities.
We may estimate all the parameters of the model by estimating each conditional probability separately by standard glm procedure. 
\\Now we turn to the misspecified model. Using the DAG structure of graph with adjacency matrices $\mathbf{A, B}$, conditionally on covariates  $\mathbf{F}$ and $\mathbf{X}$ the conditional distribution of $ \mathbf{Y} $ given symptoms $\mathbf{S}$ has the form $$P(\mathbf{Y}|\mathbf{S})=P(Y_1|Y_2,Y_3,Y_4,S_1)P(Y_2|Y_4,S_2)P(Y_3|Y_4,S_2)P(Y_4|Y_5,S_3)P(Y_5).$$
Now we are going to give specific equations restricting attention to the misspecified model only.
The first equation of our model concerns the logit for asthma $Y_1$  conditionally on symptom of asthma $S_1$, family history of allergy diseases $F_1, F_2, F_3, F_4, F_5$, and diseases $Y_2, Y_3, Y_4$, and control covariates $X_1, X_2, X_3, X_4$ 
$$logit_1=\omega_{01}+\sum_{j=1}^{4}\alpha_{j1}X_j+\sum_{j=1}^{5}\beta_{j1}F_j+\gamma_{11}S_1+\sum_{j=2}^{4}\omega_{j1}Y_j.$$
The second equation presents the logit for intermittent allergic rhinitis $Y_2$  conditionally on symptom of allergic rhinitis $S_2$, family history of allergy diseases $F_1, F_2, F_3, F_4, F_5$ and allergic dermatitis $Y_4$, and control covariates $X_1, X_2, X_3, X_4$ 
$$logit_2=\omega_{02}+\sum_{j=1}^{4}\alpha_{j2}X_j+\sum_{j=1}^{5}\beta_{j2}F_j+\gamma_{22}S_2+\omega_{42}Y_4.$$
Next, we present logit form for chronic allergic rhinitis $Y_3$ conditionally on symptom of allergic rhinitis $S_2$, allergic dermatitis $Y_4$, family history of allergy diseases $F_1, F_2, F_3, F_4, F_5$, and $Y_4$, and control covariates $X_1, X_2, X_3, X_4$ 
$$logit_3=\omega_{03}+\sum_{j=1}^{4}\alpha_{j3}X_j+\sum_{j=1}^{5}\beta_{j3}F_j+\gamma_{32}S_2+\omega_{43}Y_4.$$
Logit for allergic dermatitis $Y_4$ conditionally on symptom of allergic dermatitis $S_3$, food allergy $Y_5$, family history of allergy diseases $F_1, F_2, F_3, F_4, F_5$ and control covariates $X_1, X_2, X_3, X_4$ has the form
$$logit_4=\omega_{04}+\sum_{j=1}^{4}\alpha_{j4}X_j+\sum_{j=1}^{5}\beta_{j4}F_j+\gamma_{43}S_3+\omega_{54}Y_5.$$

In both generative and misspecified models, estimation is carried out for each equation using the standard procedure for estimating logistic regression coefficients. 
\\Next,  we compute the 'diagnostic' probabilities of diseases given symptoms for both models with estimated parameters: $P(Y_1=1|Y_2,Y_3,Y_4,S_1), P(Y_2=1|Y_4,S_2), P(Y_3=1|Y_4,S_2), P(Y_4=1|Y_5,S_3), P(Y_5)$. It is worth noting that in the case of a large network, it would not be possible to calculate $P(\mathbf{Y}|\mathbf{S,F,X})$ or $P(Y_i|\mathbf{S,F,X})$ exactly in the  generative model. In this situation, the misspecified model has an advantage over the generative model. The two methods can be compared in the case of a small network as that considered here.
\\We consider five scenarios of covariates $\mathbf{X}, \mathbf{F}$:
\begin{itemize}
	\item Case 1: rural area, children 13-14 y.o., male, without allergy history in family $F_1=\ldots=F_5=0$  and without symptoms $S_1=S_2=S_3=0$;
	\item Case 2: rural area, children 13-14 y.o., male, without allergy history in family $F_1=\ldots=F_5=0$  and with symptoms $S_1=S_2=S_3=1$;
	\item Case 3: urban area, children 13-14 y.o., male, without allergy history in family $F_1=\ldots=F_5=0$  and with symptoms $S_1=S_2=S_3=1$;
	\item Case 4: urban area, children 13-14 y.o., male, without allergy history in family $F_1=\ldots=F_5=0$  and without symptoms $S_1=S_2=S_3=0$;
	\item Case 5: urban area, children 13-14 y.o., male, with allergy history in family $F_1=\ldots=F_5=1$  and with symptoms $S_1=S_2=S_3=1$.
	
\end{itemize}
The results are presented in Tables \ref{tabX1}-\ref{tabX2}. The obtained difference between the two models is negligible.  
\begin{table}
	\caption{\label{tabX1} Results for comparison the generative model (exact computation) and misspecified model}
	\begin{center}
		\fbox{%
			\begin{tabular}{*{6}{c}}
				\em case & method & $P(Y_1|Y_2=0,Y_3=0,Y_4=0,S_1)$& $P(Y_2|Y_4=0,S_2)
				$ & $P(Y_3|Y_4=0,S_2)$ & $P(Y_4|Y_5=0,S_3)$  \\
				\hline
				Case 1&exact&0.021 &	0.081 &	0.077 &	0.024\\
				     &misspec.&0.023 &0.085 &0.080&0.015\\
				Case 2&exact&0.103 &	0.282 &	0.322 &	0.208\\
				     &misspec.&0.088 &	0.270 &	0.307 &	0.081\\
				Case 3&exact&0.074 &	0.212 &	0.321 &	0.248\\
				&misspec.&0.064 &	0.202 &	0.306 &	0.116\\
				Case 4&exact&0.015 &	0.056 &	0.074 &	0.007\\
				&misspec.&0.016 &	0.060 &	0.080 &	0.022\\     
			    Case 5&exact&0.120 &	0.332 &	0.359 &	0.233\\
				&misspec.&0.130 &	0.295 &	0.314 &	0.210\\ 		     	
		\end{tabular}}
	\end{center}
\end{table}
\begin{table}
	\caption{\label{tabX2} Results for comparison the generative model (exact) and misspecified model}
	\begin{center}
		\fbox{%
			\begin{tabular}{*{6}{c}}
					\em case & method & $P(Y_1|Y_2=1,Y_3=1,Y_4=1,S_1)$& $P(Y_2|Y_4=1,S_2)
				$ & $P(Y_3|Y_4=1,S_2)$ & $P(Y_4|Y_5=1,S_3)$  \\
				\hline
				Case 1&exact&0.597 &	0.104 &	0.134 &	0.024\\
				&misspec.&0.566 &0.097 &0.125 &	0,044\\
				Case 2&exact&0.886 &	0.326 &	0.461 &	0.524\\
				&misspec.&0.842 &	0.299 &	0.421 &	0.216\\
				Case 3&exact&0.845 &	0.250 &	0.463 &	0.581\\
				&misspec.&0.793 &	0.226 &	0.420 &	0.290\\
				Case 4&exact&0.509 &	0.073 &	0.130 &	0.029\\
				&misspec.&0.482 &	0.068 &	0.124 &	0.064\\     
				Case 5&exact&0.902 &	0.398 &	0.511 &	0.561\\
				&misspec.&0.893	& 0.326 &	0.429 &	0.452\\ 
		\end{tabular}}
	\end{center}
\end{table}

\subsection{Results of estimation for the misspecified model}

Model estimation is performed separately for each equation (see formula (\ref{ff4})-(\ref{ff5})) using the usual glm procedure for logistic regression in the first scenario, and in the second scenario, we use weighted logistic regression (with weights as in Section 2).
\\According to work of \cite{b14} we assume that the population fraction in Poland for considered allergy diseases are as follows  $\tau_1=11\%, \tau_2=20\%,\tau_3=4\%,\tau_4=7\%,\tau_5=10\%.$

\medskip

The results for estimation for two scenarios are given in Tables \ref{tab1a}-\ref{tab2b}.The standard errors for standard logistic regression coefficients estimation are given in Tables \ref{tab1ab}-\ref{tab1bb}. Next, we present the odds ratio with the asymptotic 0.95 confidence interval (CI) for the standard logistic regression and the odds ratio for weighted logistic regression (see Tables \ref{tab4a}-\ref{tab5b}).

\begin{table}
	\caption{\label{tab1a} Estimation results for standard logistic regression - Part 1}
	\begin{center}
	\fbox{%
		\begin{tabular}{*{11}{c}}
			\em $logit_i$ & $\omega_{0i}$ & $\alpha_{1i}$& $\alpha_{2i}$ & $\alpha_{3i}$ & $\alpha_{4i}$ & $\beta_{1i}$ & $\beta_{2i}$ & $\beta_{3i}$ & $\beta_{4i}$  & $\beta_{5i}$ \\
			\hline
			i=1&-5.933&0.355&0.216&-0.336&0.412&-0.075&-0.108&0.194& 0.029 & 0.736\\
			i=2&-4.000&0.287&0.364&-0.384&-0.038&-0.045&0.284&0.313&0.090 & -0.135\\
			i=3&-4.741&0.335&0.271&-0.004& 0.334&0.055&0.186&-0.039&-0.144& -0.022\\
			i=4&-6.073&0.195&-0.623& 0.395&-0.103&0.331&0.220&0.061&-0.380& 0.470\\	
		\end{tabular}}
		\end{center}
	\end{table}
	\begin{table}
		\caption{\label{tab1b} Estimation results for standard logistic regression - Part 2}
		\begin{center}
		\fbox{%
			\begin{tabular}{*{8}{c}}
				\em $logit_i$& $\gamma_{i1}$& $\gamma_{i2}$& $\gamma_{i3}$& $\omega_{2i}$& $\omega_{3i}$& $\omega_{4i}$ & $\omega_{5i}$\\
				\hline
				i=1&1.412&-&-&1.265&2.040&0.713&-\\
				i=2&-&1.379&-&-&-&0.143&-\\
				i=3&-&1.628&-&-&-&0.496&-\\
				i=4&-&-&1.780&-&-&-&1.132\\
			\end{tabular}}
			\end{center}
		\end{table}
	\begin{table}
		\caption{\label{tab2a} Estimation results for weighted logistic regression - Part 1}
	\begin{center}
	\fbox{%
	\begin{tabular}{*{11}{c}}
	\em $logit_i$ & $\omega_{0i}$ & $\alpha_{1i}$ & $\alpha_{2i}$ & $\alpha_{3i}$ & $\alpha_{4i}$ & $\beta_{1i}$ & $\beta_{2i}$ & $\beta_{3i}$ & $\beta_{4i}$ &  $\beta_{5i}$ \\
	\hline 
	i=1&-5.053&0.356&0.206&-0.349&0.447&-0.039&-0.092&0.181& 0.024&0.737\\
	i=2&-3.544&0.289&0.355&-0.377&-0.042&-0.041&0.285&0.310&0.088&-0.154\\
	i=3&-6.212&0.319&0.263&0.024& 0.341&0.061&0.186& -0.042&-0.163&0.001\\
	i=4&-3.301&0.139&-0.646& 0.278&-0.104&0.414&0.322&-0.009&-0.480&0.409\\		
	\end{tabular}}
	\end{center}
	\end{table}
\begin{table}
\caption{\label{tab2b}Estimation results for weighted logistic regression - Part 2}
\begin{center}

\fbox{%
	\begin{tabular}{*{8}{c}}
	\em $logit_i$ & $\gamma_{i1}$ & $\gamma_{i2}$ & $\gamma_{i3}$ & $\omega_{2i}$ & $\omega_{3i}$ & $\omega_{4i}$ & $\omega_{5i}$ \\
	\hline
	i=1&1.392&-&-&1.392&2.085&0.753&-\\
	i=2&-& 1.377&-&-&-&0.140&-\\
	i=3&-&1.628&-&-&-&0.450&-\\
	i=4&-&-&1.800&-&-&-&1.302\\
	\end{tabular}}
	\end{center}
\end{table}	
	
\begin{table}
	\caption{\label{tab1ab} The standard errors of estimation for standard logistic regression - Part 1}
	\begin{center}
		\fbox{%
			\begin{tabular}{*{11}{c}}
				\em $logit_i$ & $\omega_{0i}$ & $\alpha_{1i}$& $\alpha_{2i}$ & $\alpha_{3i}$ & $\alpha_{4i}$ & $\beta_{1i}$ & $\beta_{2i}$ & $\beta_{3i}$ & $\beta_{4i}$  & $\beta_{5i}$ \\
				\hline
				i=1&0.350&0.197&0.190&0.219&0.149&0.206&0.228&0.180& 0.304 & 0.304\\
				i=2&0.212&0.121&0.113&0.133&0.088&0.123&0.134&0.107 & 0.184&0.239\\
				i=3&0.223&0.116&0.110&0.142& 0.086&0.117&0.130&10.110&0.185& 0.224\\
				i=4&0.393&0.145&0.163& 0.244&0.127&0.150&0.169&0.151&0.243& 0.260\\	
		\end{tabular}}
	\end{center}
\end{table}
\begin{table}
	\caption{\label{tab1bb} The standard errors of estimation for standard logistic regression - Part 2}
	\begin{center}
		\fbox{%
			\begin{tabular}{*{8}{c}}
				\em $logit_i$& $\gamma_{i1}$& $\gamma_{i2}$& $\gamma_{i3}$& $\omega_{2i}$& $\omega_{3i}$& $\omega_{4i}$ & $\omega_{5i}$\\
				\hline
				i=1&0.151&-&-&0.184&0.158&0.222&-\\
				i=2&-&0.096&-&-&-&0.164&-\\
				i=3&-&0.096&-&-&-&0.148&-\\
				i=4&-&-&0.167&-&-&-&0.147\\
		\end{tabular}}
	\end{center}
\end{table}	
	
\begin{table}
	\caption{\label{tab4a} The OR with 0.95 CI for estimation results for standard logistic regression - Part 1}
	\begin{center}
		\fbox{%
			\begin{tabular}{*{10}{c}}
				\em $logit_i$ & $\exp(\alpha_{1i})$& $\exp(\alpha_{2i})$ & $\exp(\alpha_{3i})$ & $\exp(\alpha_{4i})$ \\
				\hline
				i=1&1.426(0.969;2.098)&1.121(0.855;1.801)&0.715(0.465;1.098)&1.510(1.127;2.022)\\
				i=2 &1.332(1.051;1.689)&1.439(1.153;1.796)&0.681(0.525;0.884)&0.963(0.810;1.144) \\
				i=3&1.398(1.114;1.755)&1.311(1.057;1.627)&0.996(0.754;1.316)&1.397(1.180;1.653)\\
				i=4&1.215(0.915;1.615)&0.536(0.390;0.738)&1.484(0.920;2.395)& 0.902(0.703;1.157)\\	
		\end{tabular}}
	\end{center}
\end{table}
\begin{table}
	\caption{\label{tab4b} The OR with 0.95 CI for estimation results for standard logistic regression - Part 2}
	\begin{center}
		\fbox{%
			\begin{tabular}{*{6}{c}}
				\em $logit_i$& $\exp(\beta_{1i})$& $\exp(\beta_{2i})$& $\exp(\beta_{3i})$& $\exp(\beta_{i4})$& $\exp(\beta_{i5})$\\
				\hline
				i=1&0.928(0.620;1.389)&0.898(0.574;1.403)&1.214(0.853;1.728)&1.029(0.567;1.868)& 2.088(1.150;3.788)\\
				i=2&0.956(0.751;1.217)&1.328(1.022;1.727)&1.368(1.109;1.687)&1.094(0.763;1.569)&0.874(0.547;1.396)\\
				i=3&1.057(0.840;1.329)&1.204(0.934;1.554)&0.962(0.775;1.193)&0.866(0.603;1.244)&0.978(0.631;1.517)\\
				i=4&1.392(1.038;1.868)&1.246(0.895;1.735)&1.063(0.791;1.429)&0.684(0.425;1.101)&1.600(0.961;2.663)\\
		\end{tabular}}
	\end{center}
\end{table}
\begin{table}
	\caption{\label{tab4b} The OR with 0.95 CI for estimation results for standard logistic regression - Part 3}
	\begin{center}
		\fbox{%
			\begin{tabular}{*{4}{c}}
				\em $logit_i$& $\exp(\gamma_{i1})$& $\exp(\gamma_{i2})$& $\exp(\gamma_{i3})$\\
				\hline
				i=1&4.104(3.053;5.518)&-&-\\
				i=2&-&4.015(3.326;4.846)&-\\
				i=3&-&5.094(4.220;6.148)&-\\
				i=4&-&-&5.930(4.275;8.226)\\
		\end{tabular}}
	\end{center}
\end{table}
\begin{table}
	\caption{\label{tab4b} The OR with 0.95 CI for estimation results for standard logistic regression - Part 4}
	\begin{center}
		\fbox{%
			\begin{tabular}{*{5}{c}}
				\em $logit_i$& $\exp(\omega_{2i})$& $\exp(\omega_{3i})$& $\exp(\omega_{4i})$ & $\exp(\omega_{5i})$\\
				\hline
				i=1&3.543(2.470;5.082)&7.691(5.642;10.482)&2.040(1.320;3.152)&-\\
				i=2&-&-&1.154(0.837;1.591)&-\\
				i=3&-&-&1.642(1.229;2.195)&-\\
				i=4&-&-&-&3.102(2.325;4.138)\\
		\end{tabular}}
	\end{center}
\end{table}
\begin{table}
	\caption{\label{tab5a} The OR for estimation results for weighted logistic regression - Part 1}
	\begin{center}
		\fbox{%
			\begin{tabular}{*{10}{c}}
				\em $logit_i$ & $\exp(\alpha_{1i})$& $\exp(\alpha_{2i})$ & $\exp(\alpha_{3i})$ & $\exp(\alpha_{4i})$ & $\exp(\beta_{1i})$ & $\exp(\beta_{2i})$ & $\exp(\beta_{3i})$ & $\exp(\beta_{4i})$  & $\exp(\beta_{5i})$ \\
					\hline
					i=1&1.428&1.229&0.705&1.564&0.962&0.912&1.198&1.024& 2.090 \\
					i=2 & 1.335&1.426&0.686&0.959&0.960&1.330&1.363&1.092&0.857 \\
					i=3&1.376&1.301&1.024&1.406& 1.063&1.204&0.959&0.850&1.001\\
					i=4&1.149&0.524&1.320& 0.901&1.513&1.380&0.991&0.619&1.505\\		
		\end{tabular}}
	\end{center}
\end{table}
\begin{table}
	\caption{\label{tab5b} The OR for estimation results for weighted logistic regression - Part 2}
	\begin{center}
		
		\fbox{%
				\begin{tabular}{*{8}{c}}
				\em $logit_i$& $\exp(\gamma_{i1})$& $\exp(\gamma_{i2})$& $\exp(\gamma_{i3})$& $\exp(\omega_{2i})$& $\exp(\omega_{3i})$& $\exp(\omega_{4i})$ & $\exp(\omega_{5i})$\\
				\hline
				i=1&4.023&-&-&4.023&8.045&2.123&-\\
				i=2&-&3.963&-&-&-&1.150&-\\
				i=3&-&5.094&-&-&-&1.568&-\\
				i=4&-&-&6.050&-&-&-&3.677\\
		\end{tabular}}
	\end{center}
\end{table}	
\subsection{Evaluation of the misspecified model}

The parameter estimator's accuracy and precision and the model's robustness for both scenarios were assessed using bootstrap and jackknife techniques. We draw 20 ordinary non-parametric bootstrap samples, calculate regression coefficients on each bootstrap sample, treat the whole real sample as a test sample, draw the ROC for it, and calculate the AUC. We also used the jackknife (10-fold cross-validation) method: we drew $10\%$ of the sample for testing and treated the other $90\%$ as a training sample. We repeated the experiment 20 times. 
Table \ref{tab3} shows the AUC values for the weighted logistic regression estimate (weight), and the averaged AUC values for bootstrap (with and without weights) and jackknife (with and without weights). 

\begin{figure}
	\includegraphics[width=9cm, height=9cm]{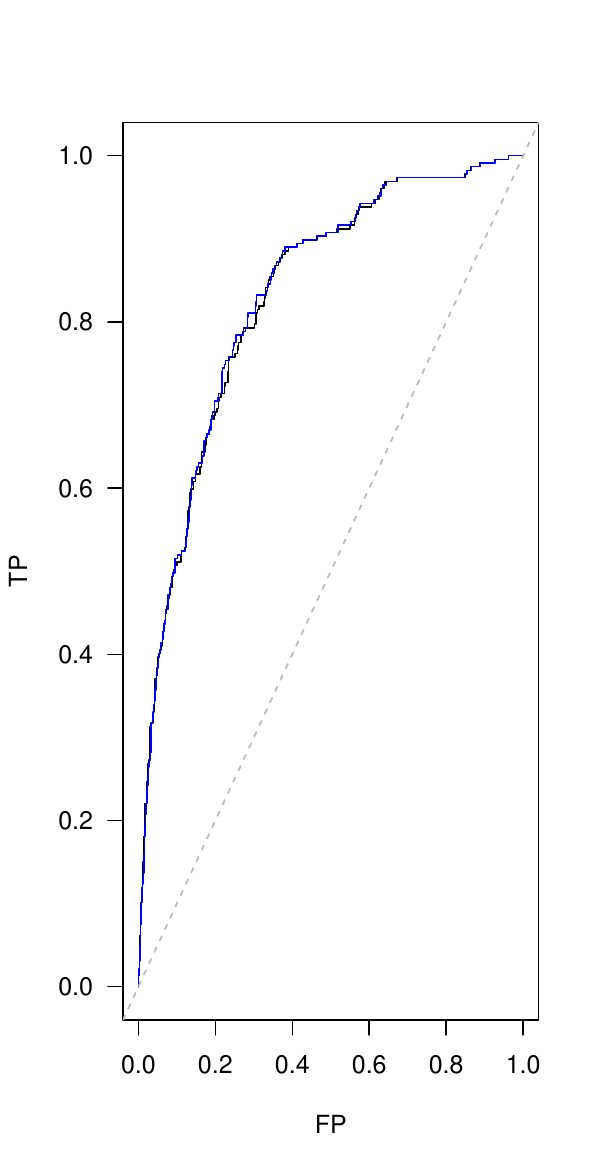}
	\caption{\label{Fig3} ROC for $logit_1$ for unweighted (black curve) and weigthed (blue curve) estimation.}
\end{figure}

\begin{figure}
	\includegraphics[width=9cm, height=9cm]{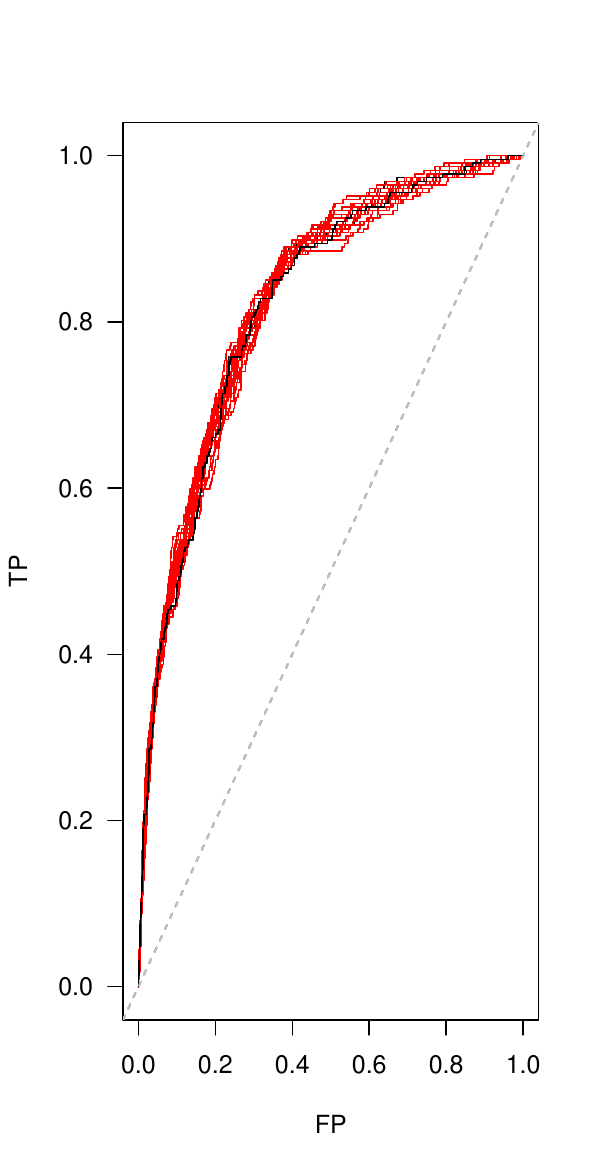}
	\caption{\label{Fig4} ROC for $logit_1$ for unweighted estimation-bootstrap.}
\end{figure}

\begin{figure}
	\includegraphics[width=9cm, height=9cm]{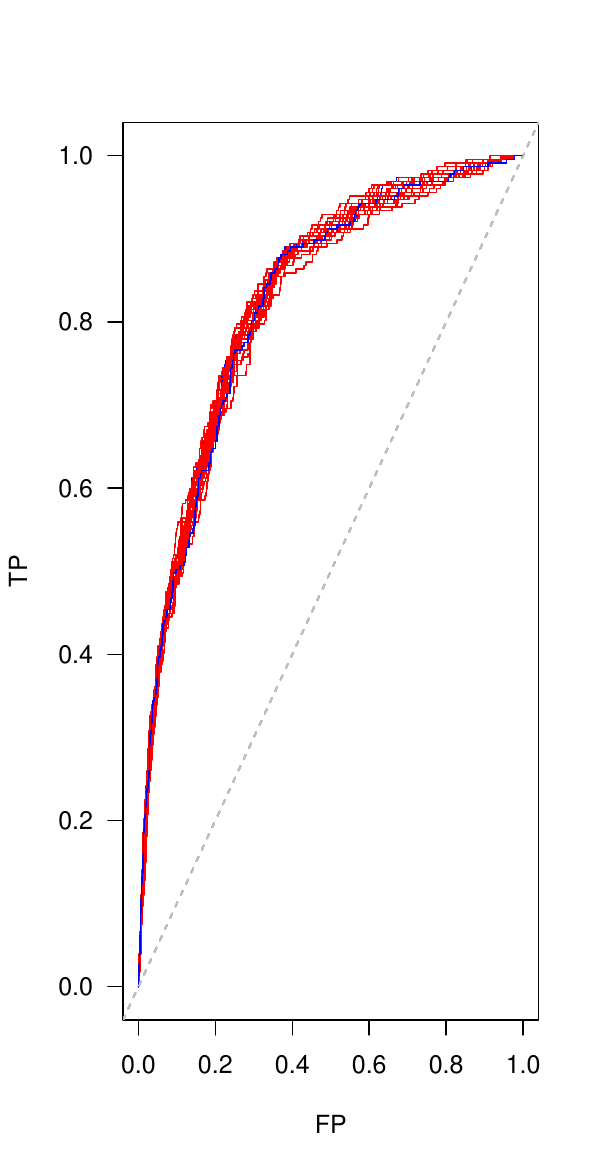}
	\caption{\label{Fig5} ROC for $logit_1$ for weighted estimation-bootstrap.}
\end{figure}

\begin{figure}
	\includegraphics[width=9cm, height=9cm]{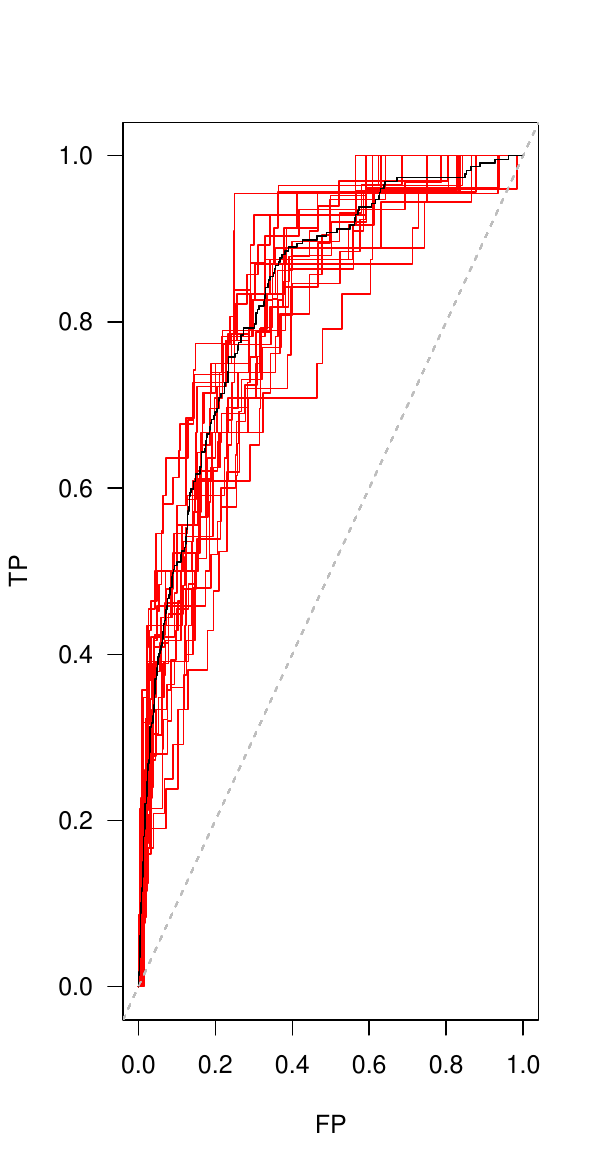}
	\caption{\label{Fig6} ROC for $logit_1$ for unweighted estimation-jackknife.}
\end{figure}

\begin{figure}
	\includegraphics[width=9cm, height=9cm]{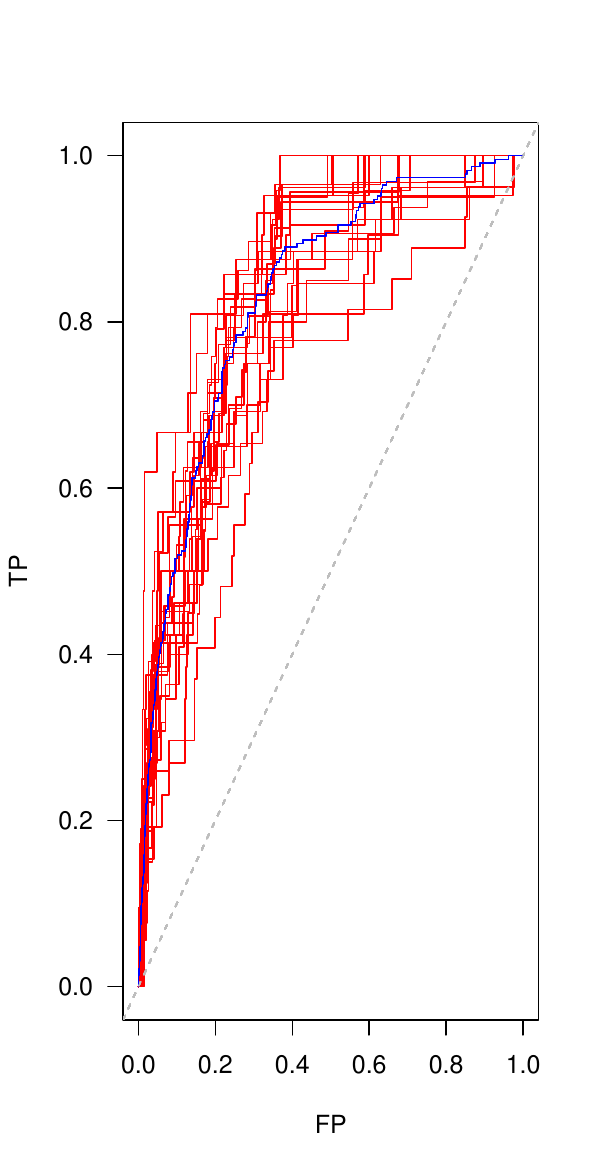}
	\caption{\label{Fig7} ROC for $logit_1$ for weigthed estimation-jackknife.}
\end{figure}

In Figure \ref{Fig3}-\ref{Fig7}, we may assess the stability and good accuracy and precision of the first logit equation. 
The rest of the results (Fig1S-Fig15S) are collected in supplementary materials on GitHub (\cite{b14a}). 
The ROC curves for the standard logistic regression and weighted estimations are almost identical except for the case of $logit_2$ (Fig 1S), where better results are obtained from the standard glm estimation method. Using bootstrap and jackknife, even for a few repetitions, showed quite good stability of the obtained results. The average value of AUC (Tab \ref{tab3}) confirmed this tendency. In general, the method without weights gave better AUC results except in the case of $logit_4$. However, the differences were quite negligible. When the jackknife method was used, also clear difference was not observed.

\begin{table}
	\caption{\label{tab3} AUC for each logit}
	\begin{center}
	\fbox{%
		\begin{tabular}{*{6}{c}}
			\em $logit_i$ &\em weight &\em boot+weight&\em boot&\em jackkn+weight&\em jackkn\\
			\hline
			i=1&0.8406&0.8258&0.8470&0.8231&0.8165\\
			i=2&0.6704& 0.6907&0.6986&0.6700&0.6857\\
			i=3&0.7234&0.7196&0.7201&0.7259&0.7215\\
			i=4&0.7971&0.7936&0.7931&0.7861&0.7921\\			
		\end{tabular}}
		\end{center}
	\end{table}	

\subsection{Results of statistical inference}

Since the differences in estimation by the two considered methods turned out to be negligible, we will make model inference based on the parameter estimation without weighting. Next, we present the most influential factors affecting each of considered disease, such as other diseases, symptoms, and other additional variables. Based on Table \ref{tab4a}-\ref{tab4b}, we may conclude that atopic asthma $Y_1$ is most associated with its symptom: wheezing or whistling in chest $S_1$ ($OR=4.104$), intermittent allergic rhinitis $Y_2$ ($OR=3.543$), chronic allergic rhinitis $Y_3$ ($OR=7.691$), grandparents on father's side has allergy $F_5$ ($OR=2.088$) and allergic dermatitis $Y_4$ ($OR=2.040$). The intermittent allergic rhinitis $Y_2$ is most associated with its symptom: problem with sneezing or a runny or blocked nose $S_2$ ($OR=4.015$) and allergic dermatitis $Y_4$ ($OR=1.154$).  The chronic allergic rhinitis $Y_3$ is most associated with its symptom: problem with sneezing or a runny or blocked nose $S_2$ ($OR=5.094$) and allergic dermatitis $Y_4$ ($OR=1.642$). The allergic dermatitis $Y_4$ is most associated with its symptom: having eczema or skin allergy $S_3$ ($OR=5.930$) and food allergy $Y_5$ ($OR=3.102$).

\section{Discussion}

Multimorbidity in allergy has been studied, among others, by \cite{b17a}, \\ \cite{b17b}, \cite{b17c}. 
\\This problem has been also considered in \cite{b18}, \cite{b19} in a cross-sectional multicentre study in Poland on the ECAP database (\cite{b14}). These studies were based on the fitting of single logistic models that did not take into account the correlations between the studied diseases. In the present approach, which is based on the conditional graphical models, we took into account such dependency through a directed acyclic graph of the relationships between the diseases under study and a directed graph of the dependency between the diseases and their symptoms. The presented model of the relationship between allergic diseases is based on discussions with medical doctors dealing with allergies and is a certain generalization of papers \cite{b18}, \cite{b19}.
The proposed model can be used in studies of associations of other diseases and, in general, in the study of correlations of complex systems. In this study, we used estimation based on a separate logistic regression estimation for the individual equations of the model as well as a weighted version of this estimation. In the case of biased data and rare diseases, we recommend using weighted logistic regression. In our case, the obtained results were very close using standard logistic and weighted logistic regressions. Due to the nature of our task, we considered the low-dimensional case where the number of observations $n$ is greater than the number of features $p$. Naturally, the proposed approach can be generalized to the high-dimensional case $p>n$ by adding the Lasso or Ridge penalty for log-likelihood for each logit model separately.

\section{Conclusions}

In our study, we presented some conditional graphical model in two versions, taking into account the known associations between allergic diseases and the symptoms of these diseases together with additional factors such as the family history of allergic diseases, and introduced additional control variables into the model. We compare two versions of our model, one in which diseases cause their symptoms (the generative model) and the misspecified model where the direction of edges is opposite. For five different scenarios of covariates, we compute the 'diagnostic' probability of diseases on given symptoms for both model versions. The obtained differences are negligible. The misspecified model approximates these diagnostic probabilities of allergic diseases very well and is less computationally expensive than calculating the exact inverse probabilities in the generative model. We will focus on the misspecified version, which we consider more practical. Parameter estimation for both versions of model were performed using the standard glm procedure for each logistic regression separately. Due to the rarity of the diseases considered, logistic regression weighting was proposed (\cite{b11}). 
Evaluation of the model using bootstrap and jackknife techniques yielded average AUCs ranging from 0.67 to 0.84 (Table \ref{tab3}), indicating fairly high stability of the results. In general, weighting did not really help in estimating model parameters, except for the equation for $logit_4$. 
\\The proposed model can be easily extended by adding other potential factors influencing the occurrence of the diseases or for general dependency complex model of binary variables. 

  \textbf{Conflict of interest}: The authors declare no conflict of interest.

\end{document}